# Improving Vulnerable Road User Safety: Existing Practices and Consideration for Using Mobile Devices for V2X Connections

Nishanthi Dasanayaka, Khondokar Fida Hasan, Charles Wang, and Yanming Feng

School of Computer Science, Science and Engineering Faculty, Queensland University of Technology, Brisbane, Australia

Corresponding author: Nishanthi Dasanayaka (email:n.mudiyanselage@qut.edu.au )

## ABSTRACT

Vulnerable road users (VRUs) such as pedestrians, cyclists, and motorcyclists are at the highest risk in the road traffic environment. Globally, over half of road traffic deaths are vulnerable road users. Substantial efforts have been made to improve VRU safety from engineering solutions to law enforcement, from the traditional Intelligent Transportation Systems (ITS) countermeasures to Cooperative-ITS (C-ITS) solutions, the death toll of VRUs continues to rise. In this paper, we start with an overview of the principles for the C-ITS approach to address the root causes of VRU safety problems and improve VRU road safety. Reduced visibility and delayed reaction time are highlighted as the most common consequences of any errors involving the driver, VRU, and other individuals in the traffic environment. C-ITS approaches enhance road safety by enabling wireless communication to exchange information among road users, such exchanged information is utilized for creating situational awareness or visibility and detecting any potential collisions in advance to take necessary measures to avoid any possible road casualties. Our studies find that the state-of-the-art solutions of C-ITS for VRU safety are limited to unidirectional communication where VRUs are only responsible for alerting their presence to drivers with the intention of avoiding collisions. This one-way interaction is substantially limiting the enormous potential of C-ITS which otherwise can be employed to devise a more effective solution for VRU safety where VRU can be equipped with bidirectional communication with full C-ITS functionalities. Based on discussions for the V2VRU communication requirements and use cases by following the C-ITS standards, the paper presents the design considerations for a smartphone-based Vehicle-to-VRU (V2VRU) communication system along with potential challenges of a Mobile Broadband (MBB) service to provide necessary recommendations.

## 1. Introduction

Owing to the absence of adequate protection and the inability to respond in crucial circumstances, road users at high risk of crash involvement are considered to be Vulnerable Road Users (VRU)[1]. This includes both non-motorized road users such as pedestrians, as well as users of VRU vehicles [2] such as bicyclists, motorcyclists, and other powered two-wheelers (i.e. electric scooters, e-bikes, etc).

Sadly, every day, somewhere in the world, many of these vulnerable road users suffer critical injuries or lose their life because of road accidents. The alarming figures from the World Health Organization in 2018 indicate the severity of the VRU safety problem as more than half of the world's road deaths are recorded amongst vulnerable road users[3]. Due to their low level of resilience and external protection, VRUs involved in road traffic accidents often sustain more severe damage than motorists. Thus, VRU crashes have become a severe global challenge over the years, both in low-income and high-income countries, perhaps due to higher vehicle usage and weakened VRU protection. For example, the Australian government statistics of road deaths in the last 10-year period from 2011-2020 indicate that almost one-third of road deaths within Australia are VRUs, and on average, there are more than two VRU related crashes per hour in Australia [4].

VRU injuries can be avoided through proactive methods that address the main causes of a crash. The majority of VRU crashes highly depend on the behavior of the driver and VRUs on the road. Additionally, problems related to the vehicles, traffic environment, and road conditions also contribute to collisions as these can create negative impacts that may interrupt the driver and other road users. Specifically, poor visibility and short reaction times can be highlighted as the most common contributors to frequent errors associated with drivers, and VRUs [5].

To protect VRUs from road accidents, governments, transportation authorities, and automotive industries have invested heavily in developing various countermeasures for VRU safety problems and VRU safety technologies. All of these existing countermeasures can be classified as either Intelligent Transportation System (ITS) or non-ITS countermeasures. The ITS countermeasures can be further classified into two groups: basic ITS and Cooperative-ITS (C-ITS). Figure 1. summarizes the major countermeasures proposed for VRU safety with the maturity of the intervention indicated.

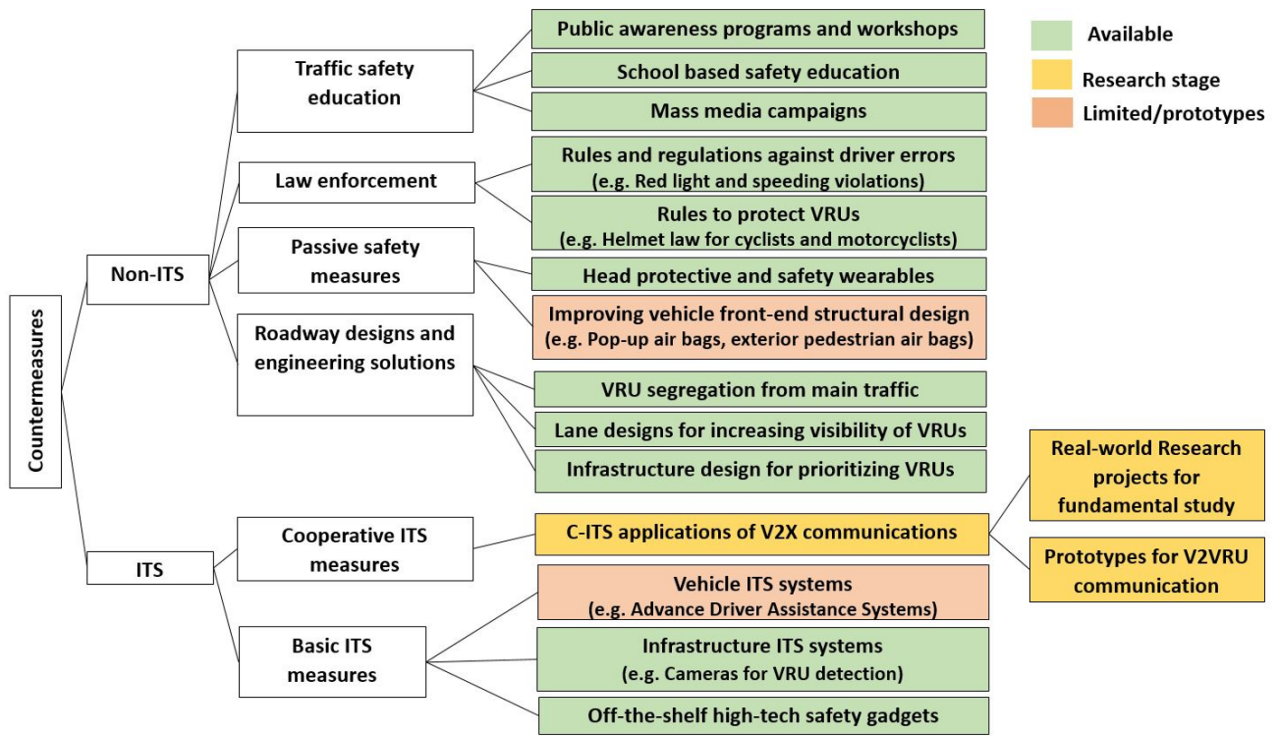

Figure 1: Overview of VRU safety countermeasures

In the past decade, significant effort has been made by governments to implement new rules and regulations to reduce driver and VRU errors, to increase awareness through safety education, and to ensure VRU protection via specific roadway design and engineering solutions. Although the above safety initiatives have avoided a considerable number of road crashes overall, there has been no significant reduction in the number of VRU-related accidents as these steps are unable to entirely resolving the root causes of crashes. For example, according to the 2019 statistical summary report of Road Trauma Australia, from 2010 to 2019, the average reduction of driver deaths is 9%, and the average reduction of passenger deaths is 27%. However, the average reduction of pedestrian and motorcyclist deaths is 6%, and pedal cyclists' deaths have increased by 3% [4]. Similarly, during the same period, hospitalized injuries of pedal cyclists' have increased by 35% [4].

Although some laws have a great potential to reduce the severity and frequency of road accidents (i.e. mandatory helmet use for cyclists and motorcyclists, restrictions on the use of mobile phones while driving), the simple fact is that in many countries where safety improvement is most needed, such regulation and enforcement has been very limited.  Similarly, there is no proper evaluation of passive safety methods that are proposed for VRU safety such as, pop-up bonnets and exterior pedestrian airbags, as most of them are still subject to reliability problems [6]. In addition to that, several studies conclude that providing dedicated bicycle-only infrastructure facilities can provide better protection to cyclists than encouraging them to wear helmets [7, 8]. Road infrastructure designs and engineering interventions, such as building bicycle-specific infrastructure facilities are straightforward and perhaps among the most effective solutions for VRU safety. They are usually implemented based on cost-benefit analysis and in the most required sections of roads, or urban areas, rather than everywhere. A large number of VRU-related accidents can be prevented by increasing awareness and interaction among cyclists and motorists.  Nevertheless, whether the infrastructure designs can create greater awareness and interaction between cyclists and motorists remains an unanswered question. Additionally, due to the challenges such as cost, feasibility, and sustainability,  experiencing infrastructure-based engineering solutions for VRU safety cannot be expected to be rapid[6].

On the other hand, leaders from academia, industry, and government have worked collaboratively to research technological solutions and thereby, accelerate the deployment of advanced technologies to tackle this global challenge. As a result, both basic and cooperative ITS applications have been introduced as a ground-breaking solution to improve traffic safety by making the road infrastructure or vehicles more intelligent.

Over the past 20 years, the focus has been on improving the car user's safety by incorporating information and communication technologies to make vehicles and infrastructure more intelligent. Advanced Driver Assistance Systems (ADAS) are an example of vehicle-based, active safety systems introduced to avoid road accidents while reducing driver errors. Various vision and radio sensors are incorporated with vehicles and roadside units to identify the presence of nearby VRUs even in non-line of sight situations to increase the visibility of VRUs to drivers and thus, alert or assist the drivers to take necessary actions within a short reaction time due to the limitation of detection distance and range.

In addition to the vehicle-ITS and infrastructure-ITS applications, several technological innovations have been recently introduced to the market for cyclists' and pedestrians' safety. The focus of such innovations is to increase the visibility of cyclists and pedestrians to motor traffic and reduce injuries by improving passive safety in case of an accident.  Helmets with advanced lighting systems to function as turn signals and brake lights [9], pair of gloves equipped with LED panel to function as turn signals [10],  and sensors enabled airbag systems [11] are few examples of high tech innovations for VRU safety.

Basic-ITS approaches mainly focus on increasing the visibility of VRUs to motor vehicles and reduce possible driver errors. However, a limitation of these applications is that they are mainly focused only on detection of the VRUs from the vehicle's viewpoint - e.g. informing the vehicle driver of the presence of the VRU. As a result, VRUs do not get a significant advantage from the basic ITS systems compared to vehicle users. More importantly, such systems do not address the weaknesses associated with VRUs that contribute to a collision. Consequently, a better approach is needed to enhance VRU safety through ITS applications that take into account the current challenges and requirements.

Cooperative Intelligent Transportation Systems (C-ITS) is another great discovery by the research community. It is an emerging technology that contributes to road safety with new vehicle technology. C-ITS enable interconnectivity among users in the traffic environment by allowing them to communicate with each other [12]. In C-ITS, road users have an important role in terms of environmental perception and information dissemination [13]. Through the enhanced interconnectivity, C-ITS approaches can address the root causes of a collision, and thereby minimize VRU related accidents. Equipped with sensors and communication technology, road users exchange up-to-date status data (i.e. location, heading, speed, etc.) with one another wirelessly to create and sustain cooperative awareness. Nearby vehicles (Vehicle-to-Vehicle, V2V), infrastructure (Vehicle-to-Infrastructure, V2I), pedestrians (Vehicle-to-Pedestrian, V2P), or other stations (Vehicle-to-Everything, V2X) can be the communication channels in this process. Significant advancements are being made in C-ITS to make V2V communications a reality in order to provide driver safety and comfort. Consequently, V2X communication systems have been mainly developed and tested for cars and trucks, showing fewer concerns for vulnerable road users and the integration of VRUs in V2X communications has been explored to a very limited extent. In such limited studies, most of the projects focus primarily on detecting VRUs from the vehicle's perspective and thereby provides collision warnings to drivers only. This leaves VRUs absent from this cooperative network and they remain as vulnerable regardless of the technological advancements. Therefore, it is crucial to integrate VRUs into the C-ITS by resolving challenges such as:

- incorporating different types of VRUs such as bicyclists and pedestrians in C-ITS,
- consideration of the most appropriate devices, communication technologies, and standards for equipping VRUs with C-ITS, and,
- assessment of challenges associated with the integration of VRUs in C-ITS.

Therefore, this paper will examine the following research questions,

1) How can C-ITS enhance VRU safety by resolving root causes of a crash including poor visibility and short reaction times?
2) What type of communication technologies, technical requirements, and use cases proposed for Vehicle-to-VRU (V2VRU) communication?
3) What are the challenges associated with the VRU integration in C-ITS using mobile devices and Mobile Broadband (MBB) service?

To address the problem, the contribution of this paper is threefold. 1). Review and appraise the current research efforts, technological solutions, and communication techniques around Cooperative-ITS (C-ITS) for VRU safety, 2). Provide design considerations for MBB based V2VRU communication outlining technological requirements, use cases, and standards to be taken into account during implementation, and 3). Discuss the challenges associate with MBB-based V2VRU implementation and make recommendations for overcoming these problems by exploring the use of new technologies along with smartphones.

This paper is structured as follows: Section II discusses how C-ITS contributes to improving VRU road safety. Section III provides an overview of state-of-art C-ITS solutions for VRU safety. Section IV discusses V2VRU communication requirements and use cases following the C-ITS standards, and provides an architecture for a MBB-based V2VRU system, identifying technical challenges and suggesting new solutions for MBB-based implementation.

## 2. Cooperative-Intelligent Transportation System (C-ITS) for VRU road safety

Enabling interconnectivity among VRUs, drivers, and infrastructure systems has the potential to be a revolutionary approach for improving VRU safety. Enabling VRUs to exchange location-specific and context awareness information between drivers with the help of wireless communication technologies offers a simple yet effective solution.

Instead of basic ITS applications that focus more on making roadside infrastructure and vehicles individually intelligent with the aid of digital technologies, integrating road users with C-ITS that use technology to enable communication between each individual in the traffic environment, will be a promising solution for enhancing VRU safety. For instance, C-ITS-equipped vehicles and roadside infrastructure have the ability to communicate a potential hazard warning to each other, allowing drivers to take the necessary actions to avoid the hazardous situation in advance. Accordingly, the C-ITS platform that initiates a cooperative, and connected transportation system will significantly improve road safety by allowing drivers to make the optimal decisions in hazardous situations.

The technology of "Vehicle-to-Everything (V2X)" communication has gained the attention of the research community for enabling connectivity in the C-ITS platform. It is proposed that a large number of road crashes can be prevented with V2X communication by allowing vehicles to effectively communicate with other individuals in the traffic environment as shown in Figure 2. Likewise, if VRUs can communicate basic state data with vehicles and infrastructure, VRUs and drivers would be alerted when their trajectories intersect. Hence, V2X technology for VRU safety would be more effective as it can mitigate collisions associated with them in advance.

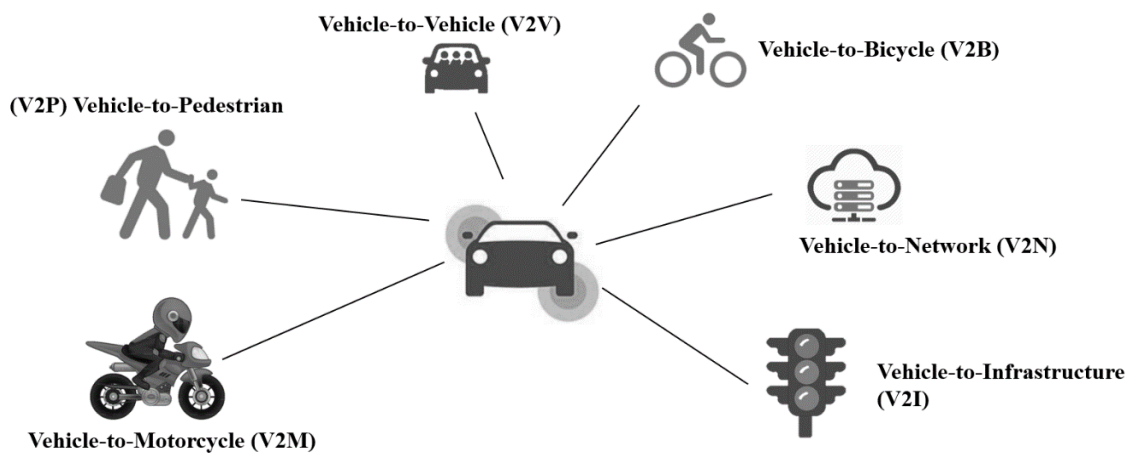

Figure 2: Overview of Vehicle-to-Everything Technology

## 3. Analysis of state-of-the-art C-ITS solutions for VRU safety

Significant research contributions and developments in the C-ITS are mainly based on V2V and V2I communications and driver-oriented road safety applications. As a result, VRU incorporation with V2X communication is a novel research area, and therefore advances around C-ITS for VRU safety are very limited. In that limited context, some research studies concern integrating VRUs into the C-ITS by discussing potential challenges related to VRU integration [14-17]. Further, several real-world research projects have been initiated to investigate and understand the requirements, use cases, and recommendations for VRU incorporation into the C-ITS platform. These include projects such as EU government-sponsored, VRUITS [18], InDEV [19], PROSPECT [20], XCYCLE [21], and projects from the Australian government including VRU and CAV (Connected and Automated Vehicle) Interactions [22], and Connected and Automated Vehicles Initiative (CAVI) vulnerable road user pilot project [23].

Besides the fundamental study of VRU integration, the academic and industry community have made significant contributions to improve VRU safety by implementing V2X communications. Several commercial and academic research prototypes are being implemented for making Vehicle-to-VRU (V2VRU) communications a reality.

Technology-based VRU safety application initiatives can be defined at two levels: awareness and collision detection, where awareness applications warn VRUs about the presence of other road users to increase the awareness of the surrounding, and collision detection applications use accurate data of shared messages to estimate their trajectories to warn about potential collision risk, and thereby, to take appropriate action to avoid the collision. Hence, such applications involve the exchange of basic safety messages between vehicles and other road users within the nearest proximity to increase awareness and alert users about hazard events.

A range of approaches is proposed, using different communication tools and technologies to communicate safety messages and exchange basic state data. These can be grouped as tag-based approaches, smartphone-based V2VRU communication approaches, and dedicated-VRU devices. The benefits and limitations of each approach are described below, and Table 1 summarizes recent research efforts of C-ITS by categorizing them based on the communication approach and the technology they have used for V2VRU communication.

Table 1: Summary of V2VRU communication systems

| VRU Device | Year | Reference | Technology | VRU type |
|---|---|---|---|---|
| Tag | 2007 | Biebl et al. [24] | RFID tag (cooperative sensors) | P, C |
| | 2008 | Fackelmeier et al.[25] | RFID transponder | P |
| | 2011 | SafeWay2School [26] | RFID tag | P |
| | 2012 | Ko-TAG [27] | RFID localization | P, C |
| | 2013 | LP3S [28] | IEEE 802.15.4 RF | P |
| Smartphone | 2012 | General Motors [29] | WiFi-Direct | P, C |
| | 2013 | Honda [30] | DSRC | P, M |
| | 2013 | Car2Pedestrian [31] | WLAN 802.11b/g/n | P, C |
| | 2014 | Wu et al. [32] | DSRC | P |
| | 2014 | WiFiHonk [33] | WiFi Beacon | P |
| | 2014 | V2ProVu [34] | IEEE 802.11g | P |
| | 2014 | Bagheri et al. [35] | Cellular | P |
| | 2016 | Lee et al. [36] | IEEE 802.11p | P |

| | 2016 | pSafety [37] | Cellular | P |
| | 2017 | WiSafe [38] | IEEE 802.11a/b/g/n | P |
| | 2017 | Artail et al. [39] | DSRC, Cellular | P |
| | 2017 | Tahmasbi et al. [40] | DSRC | P |
| | 2017 | Merdrignac et al. [41] | IEEE 802.11g | P |
| | 2017 | Telstra and Cohda wireless V2P project [42] | 4G | P, C |
| | 2018 | Zadeh et al. [43] | 4G | P |
| | 2019 | VizibleZone [44] | RFID | P |
| VRU Device V2X unit, iBeacon | 2015 | MotorWarn [45] | WiFi, DSRC, Bluetooth | C, M |
| | 2014 | Maruyama et al. [46] | DSRC | M |
| | 2017-2019 | Tome B2V app [47] | Bluetooth | C |

\* P=Pedestrian, C= Cyclist, M=Motorcyclist

In tag-based approaches, limited information is communicated between vehicles and VRUs with the aid of transmitters and receivers using technologies like Radio Frequency Identification (RFID). Such transmitters/receivers can be tagged to VRU wearables and vehicle accessories. Unlike infrared, radar, or vision-based VRU detection, tag-based approaches can function even in the Non-Line-Of-Sight (NLOS) situations. The SafeWay2School project is an example of RFID-based VRU protection which is especially proposed for providing safe and secure transportation to school children through communication with intelligent bus stops that warn drivers about the children within the proximity [26]. However, the communication radius is small and can only transmit small-sized messages such as ID-code [14]. For example, RFID-based VRU protection systems developed by Biebl et.al. [24], Fackelmeier et.al. [25], and LP$^3$S system by Lewandowski et.al. [28] could only achieve detection distance up to 80 m. A further limitation of tag-based approaches is that they cannot provide bi-directional communication where VRUs have active participation. Moreover, since tag-based approaches allow limited information to be communicated, such applications can not comply with C-ITS message standards such as Cooperative Awareness Messages (CAMs) or Decentralized Environmental Notification Messages (DENMs) and that limits the implementation of this approach to awareness applications. For example, in the LP$^3$S [28], a static short message e.g. ``Hello" is sent from the vehicle, and ``Here I am" is the response from the VRU. Thus, due to the lack of adequate data, the implementation of advanced collision avoidance applications with complex computations and filtering processes is challenging for tag-based approaches. Specifically, the distance and angle between vehicles and pedestrians are calculated based on the roundtrip time of the sent and received radio signals by the vehicle and VRU tags. Hence, the distance errors may grow due to impacts of components in the transceivers, signal noise, attenuation, and different signal propagation paths [27]. Thus, the collision avoidance computations are subject to reliability issues.

Smartphone-based approaches use applications with the aid of short or wide-range wireless communication for V2VRU situational awareness and collision detection. Tahmasbi et al.[48] propose a Dedicated Short Range Communications (DSRC) based cooperative Vehicle-to-Pedestrian (V2P) communication system, while Honda has demonstrated V2P and Vehicle-to-Motorcycle (V2M) safety applications using mobile GPS and DSRC enabled smartphones [30]. In 1999, Federal Communication Commission (FCC) proposed DSRC technology by allocating 75 MHz of bandwidth in the 5.9 GHz frequency to be used by ITS applications [49]. In addition to V2X communication, certain DSRC allocated spectrum bands are used for electronic toll collection applications in various parts of the world. The term "short-range" indicates the coverage of hundreds of meters. Comparatively, a DSRC system consisting of a roadside unit (RSU) and an on-board unit (OBU) for direct communication is known to be the fastest mode of communication. Even under high vehicle mobility conditions, DSRC technology has the potential to establish direct communication between vehicles and VRUs, and guarantee low latency of message exchange [50]. However, the main technical challenge with DSRC technology is that it requires enabling a large number of devices with DSRC capability, and vehicles should be equipped with additional DSRC equipment that is problematic for older or regular models of vehicles with no DSRC unit [33].

Further, DSRC based applications require data processing and safety warning computation at the user end which requires smartphones with high computing power [50]. Additionally, such approaches create scalability problems where the communication channel can easily get congested when multiple devices actively participate in the communication within close proximity [48, 51].

Besides DSRC based communication, there are several developments around Wireless Local Area Network (WLAN) for V2VRU communication. For example, General Motors [29] and Lee et.al. [36] have proposed a pedestrian detection technology and V2P communication system using Wi-Fi Direct to establish an ad-hoc network between the smart devices in vehicles and smartphones carried by VRUs. WiFiHonk by Dhondge et al. [33] is another example of a WLAN smartphone-based Car2X communication application that provides warnings of possible collisions between the VRU and vehicle using beacon stuffed WiFi communication. Similarly, Car2Pedestrian communication [31], V2ProVu [34], and WiSafe [38] are more examples of WiFi-based smartphone applications that use GPS sensor data for trajectory prediction and collision detection. The key limitation of WLAN systems is the limited communication range and increased communication overhead due to channel congestion and connection handovers. Based on the experimental results of WiFi-based systems, it is noteworthy that, packet loss rate increases with increases in vehicle mobility, speed, distance, and when two vehicles are crossing each other at high speeds [29],[33],[34]. The experiment results of the V2ProVu system show that to achieve a packet delivery rate of 80%, the distance needs to be smaller than 130 m [34]. If the WiFi signals are obstructed by anything such as the human body, trees, vehicles, etc. the communication distance would be significantly shorter [34]. For example, Engel et.al. [52], have achieved a 200m communication range in case of no obstacles, however, in a crowded parking area, the range has been reduced to approximately 60 m. Hence, the obtained communication range is not sufficient to transmit early warnings at high mobility since the vehicles will be connected to the Wi-Fi network only for a very short time reducing the time available for the driver's response to the collision warning [52]. Thus, the WLAN communication range decreases due to physical factors such as transmitting power, antenna gain, the geometry of the vehicle, and environmental barriers. Therefore, the performance of such systems depends on the vehicle speed, the distance between two users, and the number of users per network.

Cellular technology is also a strong candidate for implementing VRU safety applications due to the fact that it is already available for consumer devices. In recent years, cellular technologies such as LTE have been used with smartphones for implementing V2VRU applications to provide wide-ranged communication between vehicles and VRUs. V2P collision avoidance application by Bagheri et al. [35] is one good example of cellular-based V2P communication. Nevertheless, cellular-based communication involves packet routing through base stations that may increase communication latency compared to direct communication such as DSRC based communication.

To overcome the key limitations of each communication technology, hybrid approaches based on the combination of technologies have been investigated for better performance. In fact, hybrid solutions combining cellular technologies and ad-hoc networks have been proposed from previous work. For example, pedestrian safety systems by Artail et al.[39] uses DSRC and cellular technology for establishing communication links between vehicle, smartphone, and control server.

Other VRU safety approaches are based on dedicated devices for establishing communication between vehicles and VRUs. The ``MotoWarn" system by Anaya et al. [45] uses Wi-Fi, DSRC, and Bluetooth with iBeacon to create awareness among vehicles and cyclists by informing the presence of cyclists in the proximity. Similarly, Drive C2X project by Honda develops a prototype motorcycle with C-ITS hardware that can communicate with vehicles within proximity to avoid potential collisions [46]. The tome software company develops a VRU device [47] that enables bicycle-to-vehicle communication using Bluetooth 5 technology. However, the main disadvantage of such applications is that they require the development of specific devices for communication whereas multi-capable smartphones are common among VRUs.

A V2VRU safety application should consider development and safety criterion during the implementation of the system to offer maximum benefits to its end users. Therefore, five main criteria that are important to consider during the implementation of the V2VRU system are discussed below.

1. **Cost efficiency:**
   To introduce a V2VRU system to the community, it should be cost-efficient. Thus a V2VRU system should minimize initial implementation cost including hardware and technology, and maintenance cost.
2. **Bidirectional communication**
   To provide maximum protection to VRUs from road accidents, a V2VRU system should allow VRUs to involve in V2VRU communication and safety warnings should be issued to both VRUs and drivers.
3. **Communication range**
   The communication range between the VRUs, vehicles, and infrastructure should be sufficient to estimate collision risks based on the exchanged location data.
4. **C-ITS standards and requirements**
   A V2VRU system should meet the required quality by following the defined standards and requirements. The C-ITS standards include message standards, security, and privacy standards. The requirements include positioning accuracy, communication latency, communication range, etc.
5. **Collision avoidance**
   A V2VRU system should be able to identify possible collision risks and take necessary actions to avoid the collision beforehand.

The V2VRU systems given in Table 1 are compared taking into account whether the above parameters were considered by the system during the implementation. The comparison results are shown in Table 2.

Table 2: Comparison of existing systems

| VRU device | Reference | 1 | 2 | 3 | 4 | 5 |
|---|---|---|---|---|---|---|
| Tag | Biebl et al. [24] | ✓ | ✗ | <100m | ✗ | ✗ |
|  | Fackelmeier et al.[25] | ✓ | ✗ | <100m | ✗ | ✗ |

| | | 1 | 2 | 3 | 4 | 5 |
|---|---|---|---|---|---|---|
| | SafeWay2School [26] | ✓ | ✗ | - | ✗ | ✗ |
| | Ko-TAG [27] | ✓ | ✗ | - | ✗ | ✗ |
| | LP3S [28] | ✓ | ✗ | <100m | ✗ | ✗ |
| Smartphone | General Motors [29] | ✓ | ✗ | Approx. 200m | - | ✗ |
| | Honda [30] | ✗ | ✓ | <1000m | ✓ | ✗ |
| | Car2Pedestrian [31] | ✓ | ✓ | Max 200m | ✗ | ✓ |
| | Wu et al. [32] | ✗ | ✓ | - | - | ✓ |
| | WiFiHonk [33] | ✓ | ✓ | <300m | ✗ | ✓ |
| | V2ProVu [34] | ✓ | ✗ | <200m | ✓ | ✓ |
| | Bagheri et al. [35] | - | ✗ | >1000m | - | ✓ |
| | Lee et al. [36] | ✗ | - | - | - | ✗ |
| | pSafety [37] | - | ✓ | >1000m | - | ✗ |
| | WiSafe [38] | ✓ | ✗ | 100m-600m | ✗ | ✗ |
| | Artail et al. [39] | - | ✗ | - | - | ✓ |
| | Tahmasbi et al. [40] | ✗ | ✓ | <300m | - | ✓ |
| | Merdrignac et al. [41] | - | ✓ | - | - | ✓ |
| | Telstra and Cohda wireless V2P project [42] | ✓ | ✓ | >1000m | - | ✗ |
| | Zadeh et al. [43] | - | ✓ | >1000m | - | ✓ |
| | VizibleZone [44] | - | ✗ | Approx. 150m | ✗ | ✗ |
| VRU Device V2X unit, iBeacon | MotorWarn [45] | ✗ | ✗ | <100m | ✓ | ✗ |
| | Maruyama et al. [46] | ✗ | ✗ | - | - | ✓ |
| | Tome B2V app [47] | ✓ | ✗ | <300m | ✓ | ✓ |

* 1: Cost efficiency, 2: Bidirectional communication, 3: Communication range, 4: C-ITS standards and requirements, 5: Collision avoidance

According to the comparison of Table 2, tag base approaches appear to be cost-effective relative to other approaches, as they require low-cost hardware and technology for the implementation. However, it is not a successful candidate for V2VRU applications due to its primary drawbacks such as limited communication range, one-way communication, incompatibility with C-ITS standards and requirements, and inability to support collision avoidance applications. During the implementation of the systems, many smartphone-based approaches have considered cost-efficiency criteria by selecting low-cost and available infrastructure. However, some smartphone approaches require custom smartphones to enable V2VRU communication. For example, the V2VRU systems by Honda [30], Wu et.al [32], Lee et al[36], and Tahmasbi et al. [48] need DSRC enabled smartphones for communication that are not popular among users. Unlike tag-based approaches, smartphone-based approaches provide a wider range of communication and can support collision avoidance applications. However, there is no proof that C-ITS standards are used in many smartphone-based approaches. V2VRU systems requiring special or dedicated devices for communication do not tend to be very cost-effective as they require additional costs for initial deployment, infrastructure, and maintenance. Also, they only provide safety warnings for either vehicle drivers or VRU, and not for both.

Taking into account the pros and cons of current approaches, the implementation of C-ITS applications by wireless communication and utilizing existing infrastructure and devices would be a more productive solution as it minimizes deployment costs, maintenance costs, and simplifies user acceptance. A smartphone with the mobile broadband connections such as 4G MBB and 5G MBB soon could be leveraged for V2VRU communications [53]. In fact, it could significantly reduce deployment costs, resolve reliability issues of short-range communication technologies, accelerate the market penetration of VRU communication systems, and incorporate VRUs into C-ITS allowing bi-directional communication between VRUs and vehicles. Although the mobile broadband services are not designed for vehicular communication as DSRC technology, their high mobility support, long-range communication, high bit-rate, and greater bandwidth, show the potential for V2VRU communication [54-56].

Due to the limitations of the existing efforts, there is a large gap in the real-world between what C-ITS can offer for VRU safety and what the current research efforts have focused on. In particular, most of the systems focus on VRU detection and providing awareness warnings to drivers rather than provide VRUs with any warnings related to potential collision risks of vehicles. This may be because the driver is more responsible for avoiding collisions than the VRUs since even moderate-speed vehicles contribute more energy to a crash than the VRU. Nevertheless, as a result of this unidirectional communication, the VRUs are unaware of the collision danger and the whole agency for avoiding the collision lies entirely with the driver.

Thus, it is unclear whether these so-called V2VRU communication systems have fully realized the potential road safety benefits for VRUs. Similarly, such systems are limited to awareness applications rather than implementing collision avoidance applications. Furthermore, a major issue with the existing C-ITS implementations is that there is no evidence for the consideration of V2VRU communication standards. As a result, such systems are not interoperable with heterogeneous systems.

More importantly, V2VRU communication requirements have not been considered in the implementation of many systems. For example, the positioning inaccuracy of the smartphone is not considered in most of the smartphone-based approaches [29, 30, 34, 38]. Hence, such systems are subject to reliability issues as many standard smartphones provide 5-10 m positioning accuracy. Therefore, given the fact that VRU movement patterns, response times, and crash scenarios are fundamentally different from those of vehicles [40], it is necessary to consider use cases and requirements of V2VRU communication in accordance with standards specific to VRU communication.

## 4. Design Considerations for smartphone and mobile broadband-based V2VRU safety system

As discussed above, the utilization of smartphones equipped with cellular technology for C-ITS implementations can be a successful approach in developing cost-effective V2VRU safety applications for incorporating VRUs as active players. Therefore, in this section, we provide details on design considerations for a V2VRU safety system that can be implemented using a smartphone and mobile broadband technology. In fact, we first discuss the V2VRU communication requirements and use cases; secondly, we discuss the technical opportunities of using a smartphone for V2VRU communication by considering the VRU use cases and functionalities; thirdly we discuss how the mobile broadband services can meet the V2VRU communication requirements and finally, we propose a new V2VRU system framework based on smartphone and mobile broadband technology by offering a general system architecture.

## 4.1 V2VRU Communication Requirements and Use Cases

The development of an efficient V2VRU communication system requires meeting a set of requirements; thus, it is necessary to understand the technical and application requirements for implementing a better approach for VRU safety. In the following, key requirements and use cases for V2VRU communication are discussed in compliance with the ETSI standards of VRU safety [2, 57].

### 4.1.1 Requirements for V2VRU communication

As described in the previous sections, VRU safety applications can be broadly defined at two levels, with the simplest level being 'awareness' and 'collision avoidance' the advanced level [14]. Awareness applications provide basic safety notifications to inform the presence of other road users to maintain cooperative awareness. Such applications do not require high accuracy of VRU positioning and speed, but they require a periodic broadcast of basic status data of vehicles and VRUs among each other. Conversely, collision avoidance applications provide collision risk warnings by calculating trajectories of road users; therefore, such applications require high accuracy of positioning and accurate status data such as heading direction, and speed. Given these differences, understanding the major requirements for implementing a V2VRU system is highly important. To this end, we have identified 8 parameters as the major requirements for a successful VRU communication system. These are communication range, positioning accuracy, context-awareness, communication latency, scalability, user interface and warning message design, message standardization, and security and privacy.

**Communication range**

The desired benefit of V2VRU communication is detecting hazardous situations prior to visual contact. However, it must be detected in time to avoid conflicts through early warnings and precautions. The timing of the warning depends on the Time-To-Collision (TTC) and should take into account the user reaction time, communication latency, the time required for maneuver, and a safety margin [14]. Thus, the range should be sufficient to perform a risk assessment based on the awareness messages prior to issuing the warning. The communication range required for a VRU safety application has been defined as shown in Table 1 by considering communication modes such as V2I, V2V, VRU2V, and collision avoidance purposes of VRU safety application.

**Positioning accuracy**

In a V2VRU application, the majority of safety warnings are based on the proximity of the road users; therefore, precise positioning of the user location is essential. For VRU applications, positioning accuracy requirements is significantly higher than for conventional C-ITS applications. In fact, the accuracy of current smartphone positioning systems needs to be further enhanced to achieve a centimeter or decimetre-level precise positioning to offer a reliable source of localization. According to European standards, vehicle applications require a positioning accuracy of 1m [58]. However, significantly higher precision and accuracy of positioning information are required for typical VRU use cases in order to identify whether the VRU is in a safe area or not [57]. Therefore, the precision of 0.5m or higher is defined as the required positioning accuracy for VRU safety applications [57].

**Context-awareness**

The main characteristic of a collision-avoidance system is the ability to predict movements (trajectories and momentum) with the ability to act on time (changing trajectory/reducing velocity) to avoid the collision [2, 57]. Therefore, the context of the VRU and the transition of the VRU object state (i.e. walking, cycling, standing, etc.) should be determined through such applications with the use of sensors on the VRU devices [14]. Hence, the accuracy of movement prediction should be sufficiently high to minimize the miscalculation of a conflict.

**Latency**

The end-to-end latency of data communication is a key parameter that should be minimized as it impacts the accuracy of the received data elements. In particular, the shared data should be timely enough to be useful to the receiver for the collision avoidance process, leading to a minimum end-to-end latency time (e.g. less than 300 ms) and to a sufficient data sampling rate (e.g. 10 Hz) [57].

**Scalability**

The collision avoidance applications or awareness applications should perform well with multiple numbers of road users (up to 5000 users within the same communication zone, i.e. within a circle of radius up to 300 m as defined in ETSI 103-300-2 [57]. A VRU system can achieve this by using an effective clustering approach to cluster active users in a geographical area.

**User interface and warning message design**

The user interfaces of VRU safety applications should be designed to support good decision making and timely responses of road users by considering key areas such as clear and straightforward information delivery, minimum distraction, and reduced contents and workload [59]. Safety warnings should be designed to elicit the desired reaction by road users without distracting them by sending frequent low-risk warnings or false alerts. Optimal timing for sending warning messages and warning modes for VRUs and drivers should be determined based on the VRU type and danger level of the situation.

**Message Standardisation**

Standardization of messages in V2X communication systems is highly required to enable interoperability [59]. For that reason, messaging standards for V2X communication have been defined by different authorities, such as the European Telecommunications Standards Institute (i.e. ETSI standards) and Society for Automotive Engineering (i.e. SAE standards). Nevertheless, the implementation of standards for VRU communication has been initiated most recently with case studies of VRU use cases and standardization perspectives. In particular, the latest ETSI standard (ETSI 103-300-2) [57] introduces a standard message for V2VRU communication, named VAM (VRU Awareness Message) that is different and more flexible than Cooperative Awareness Message (CAM) standard due to the shortened length and VRU specific content. More importantly, the VAM message tentatively harmonized with the Personal Safety Messages (PSM) which is the standard message defined in SAE J2735 for VRU safety communication [60]. Hence, VAM messages including VRU basic status data such as location, VRU type, speed, direction, orientation, are periodically broadcasted via VRU devices to the other road users in the system to create awareness. Alternatively, Decentralized Environmental Notification Messages (DENMs) that can be used to signal the danger of a crash involving VRU are event-driven messages that only trigger when a warning is required to inform users of a hazardous event. Therefore, the DENM message includes the event type, event location, and other information that describes the severity of the event.

**Message Size and message transmitting frequency**

A VRU safety application should be able to communicate periodic broadcast messages at a maximum frequency of 10Hz, with message payloads of 50-300 bytes, not including security-related components for cooperative awareness use cases [61]. To avoid crashes involving VRUs, such applications should be able to communicate event-triggered messages with message payload which can be up to 1200 bytes not including security-related components [61].

**Security and Privacy**

There are security and privacy concerns with VRU applications, creating the requirement for strategies to mitigate such issues. In fact, security concerns include problems regarding false positives and false negatives.

- A false positive means that a receiver thinks there is a situation that requires reaction when such a situation does not actually exist. This happens when a receiver believes a message in the VRU system is true while the message is actually false. Such a situation may negatively affect system users as it can lead the receiver to trigger an action in the real world. For example, if a false VRU message gives a driver the incorrect impression that a child was running in front of the car, that may lead to a rear-end collision as the driver suddenly hits the brakes [2].
- A false negative means that a receiver does not think there is a situation where a reaction is needed when such a situation actually occurs. This happens either when the relevant warning is not received by the receiver or, if the receiver receives the warning message, they may also receive contradictory messages which lead to disbelief in the original message. For example, a denial of service (DoS) attack might lead to a receiver not receiving any messages from VRUs [62].

Therefore, to avoid false positives, cryptographic protection for messages must be included in VRU communications using credentials issued only to trusted devices [2]. Nevertheless, it is hard to provide protection against false negatives through communication security mechanisms alone. For example, a DoS attack is unavoidable, however, such an attack can theoretically be identified, and the authorities alerted in order to physically remove the source of the attack. Also, communication security mechanisms can prevent an attack based on contradictory messages by making it harder for an invalid sender to generate convincing contradictory messages.

Additionally, VRU applications may create privacy concerns as they produce data about VRUs and other road users in the traffic environment, including personal data. Therefore, the strategies to mitigate privacy concerns should include technical measures to protect road users' data. This may include restrictions for including critical personal data in the message or changing temporary sender identifiers periodically. Moreover, data management policies on retention and access to data generated by VRU applications should be incorporated with the privacy concern mitigation strategies.

Table 3: Summary of the technical requirements for V2VRU communication system[14, 57, 58, 61, 63]

| Basic Requirement | Required Value |
|---|---|
| Communication range | >= 25 m range when VRU-to-infrastructure communication for VRU protection purpose |
| | >= 75m range when VRU- to- vehicle communication for pedestrian collision avoidance purpose (stationary pedestrian and vehicle speed at 45 km/h) |
| | >= 150m when VRU-to-vehicle communication for cyclists' collision avoidance purpose (cyclist |

| | speed at 30 km/h and vehicle speed at 90 km/h) |
|---|---|
| | >= 300m when VRU-to-vehicle communication for motorcycle collision avoidance purpose |
| Positioning accuracy | A precision of 0.5 m or higher is required |
| Context-awareness | The accuracy of VRUs' movement prediction should be high enough to minimize the miscalculation of a conflict |
| Latency | Best, less than 100ms or at least, not exceeding 300 ms |
| Scalability | Should accommodate urban scenarios with use cases that include up to 5000 users per intersection |
| User interface and warning message design | UI should be designed with minimum distraction straight forward information delivery, and reduced contents and workload |
| Message standardisation | ETSI standards, <br> 1. CAM for vehicle status communication <br> 2. VAM for VRU status communication <br> 3. DENM for event-driven messages <br> SAE standards, <br> 1. PSM for VRUs' basic status communication <br> 2. BSM for vehicle status communication <br> Security standards, <br> 1. IEEE 1609[64]: Implicit and Explicit security certificates <br> 2. ETSI ITS security certificates[63] |
| Message size (between two devices) <br> 1. Periodic broadcast messages <br> 2. Event-triggered messages | 1. 50-300 bytes, not including security-related message components <br> 2. Up to 1200 bytes not including security-related message components. |
| Message transmitting frequency | A maximum frequency of 10 messages per second (i.e. 10 Hz) per transmitting UE |
| Security and privacy | To avoid false positives, cryptographic protection for messages must be included |
| | To avoid false negatives, the attack should be identified and remove the source of the attack physically |
| | To avoid privacy issues, critical personal data should not be included in messages and temporary sender identifiers should be changed periodically |

Table 3 summarizes the technical requirements that are described from the above sub-sections for the V2VRU communication system based on the ETSI standards.

### 4.1.2 VRU Safety Use Cases and Scenarios

Use cases for VRU safety can be categorized based on the interactions/communications between VRUs, vehicles, and road infrastructure. Six categories of VRU use cases have been defined in the ETSI TR 103 300-1 standard report on VRU awareness [2]. The key categories of VRU use cases are presented in Table 4.

Table 4: Categories of VRU use cases and examples[2]

| Use case Category | Description |
| --- | --- |
| A | VRU-to-VRU direct communication<br>VRUs are equipped with a communication device |
| B | VRU-to-Vehicle direct communication<br>Vehicle and VRU both are equipped with a communication device |
| C | Communication via a third party (i.e. another vehicle)<br>A third-party vehicle communicates with other vehicles to inform the detection of a hidden VRU. The VRU is not equipped with a communication device. |
| D | Communication via a third party (i.e. infrastructure)<br>A cloud server/control centre monitoring the evaluation of VRUs via roadside equipment or VRU devices and alerting to vehicles. VRUs, vehicles, and roadside equipment equipped with communication devices. |
| E | Communication via a third party (i.e. local or cloud server)<br>A cloud server/control center monitoring the evaluation of VRUs via roadside equipment or VRU devices and alerting to vehicles. VRUs, vehicles, and roadside equipment equipped with communication devices. |
| F | Communication via a third party (i.e. roadside unit)<br>Roadside equipment monitoring the evaluation of VRU via VRU communication detecting the risk of collision and alerting to approaching vehicles. VRUs, vehicles, and roadside equipment equipped with communication devices |

The categories are defined to implement with short-ranged communication technologies with dedicated infrastructure such as V2X units for road users and Road-Side-Units (RSU). However, there are differences in the way a smartphone and cellular-based system can address the given use case categories to the short-range communication system. In fact, many cellular-based systems have considered the use cases that fall under category E, which is communication via the assistance of a third party (control server/center). Figure 3-Figure 6 are a few examples of use cases of category E of cellular-based systems that have been considered in the literature.

It is notable that most of the cellular-based systems adopt a client-server architecture where client devices (i.e. smartphones) provide basic data related to VRUs and vehicles while the server performs collision predictions and send warnings back to users. However, not being limited to the use case examples of Figure 3-Figure 6, many use case scenarios presented by Scholliers et al.[14] and ETSI TR 103 300-1 report [2], and that comes under the other five categories, can also be implemented with a cellular-based approach by making necessary changes. In fact, the use case examples of road-sharing between VRUs and vehicles, turning vehicles while VRU approaching, and a VRU crossing road at crosswalks or a location remote to a crosswalk can be implemented with a cellular-based approach using a smartphone as a user device and control server or infrastructure as a communication middleware.

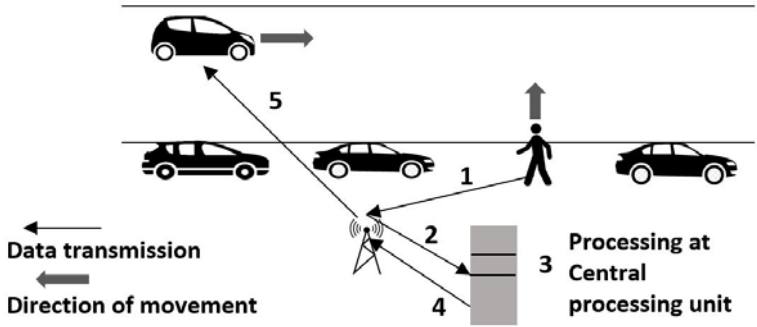

Figure 3: VRU crossing behind the parked cars at NLOS situation, Car-2-X pedestrian safety system[65]

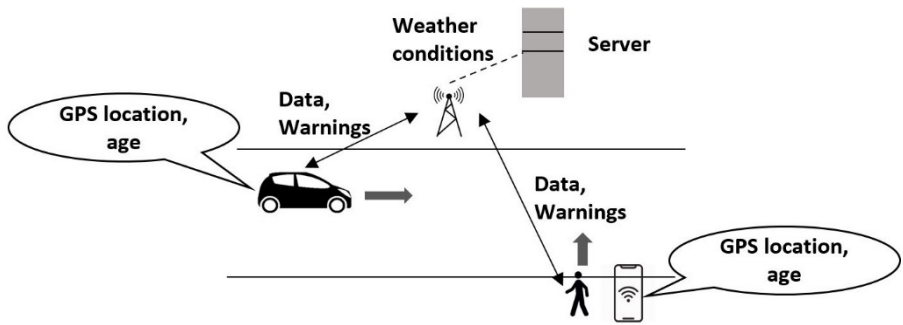

Figure 4: VRU crossing when a vehicle approaching at LOS situation, warning system by Zadeh et al.[43]

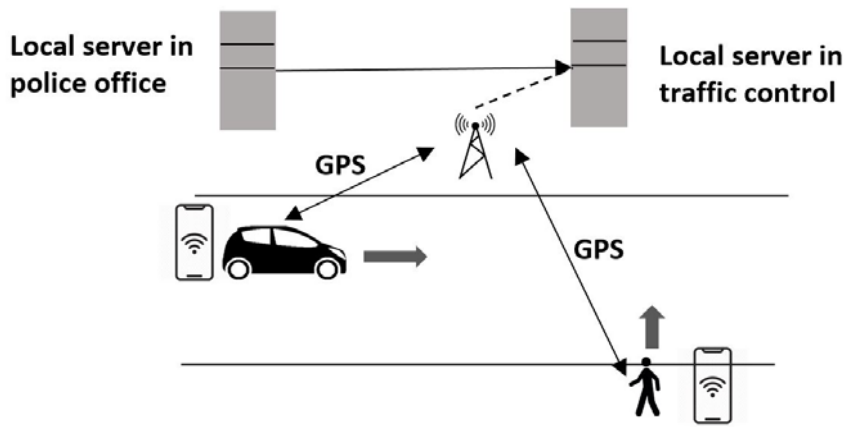

Figure 5: Warn vehicles of distracted pedestrians at both NLOS and LOS situations, pSafety system[37]

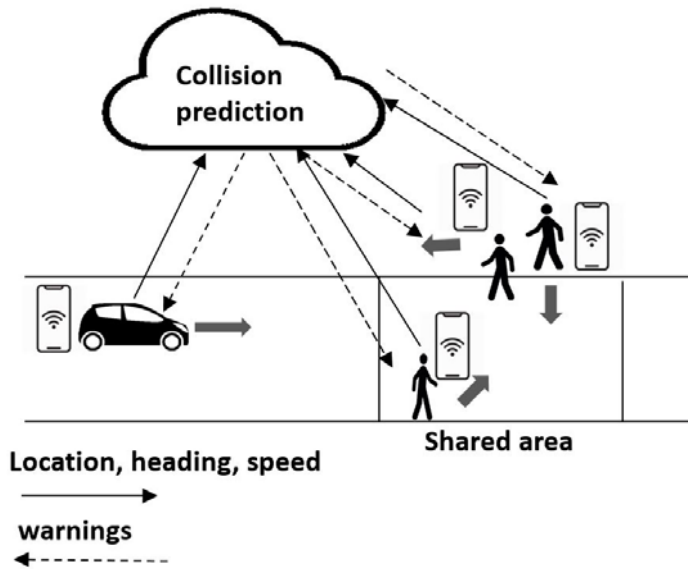

Figure 6: Collision warnings for both vehicle and pedestrians in situations such as sharing the road, and crossing road situations[35]

### 4.2 Technical opportunities of using smartphones for V2VRU communication

There are key advantages of using smartphones for V2VRU communication. First, mobile devices including smartphones, and tablets have been considered as one of ITS sub-systems in the global communication architecture for Intelligent Transport System Communication (ITSC) since its inception. For instance, the personal ITS sub-system (i.e. in hand-held devices such as mobile phones) is one of four ITS sub-systems specified in the European standards, "ETSI EN 302 665 V1.1.1 (2010-09): Intelligent Transport Systems (ITS); Communication Architecture" [66]. According to the ESTI TR 103 300-1 for VRU awareness (2019-09) [2], a VRU system is specified as an ensemble of ITS stations interacting with each other to support VRU use cases. In fact, smartphones are considered as an ITS-Station in a nomadic ITS sub-system in the context of a portable device, that can support VRU use cases and interact with vehicles, infrastructure, and other entities in the V2X network. The primary advantage of using a smartphone as a personal ITS-station is the portable characteristics of the device. This is because, for the system it contributes to time-variant roles for the device. For instance, a smartphone can be used as a VRU device for a cyclist while riding a bicycle, and it will soon become a pedestrian VRU device when the person gets off the bicycle. Additionally, VRU system is a heterogeneous system comprising a variety of VRU profiles in which a VRU can be a transmitter where a VRU equipped with ITS-station may only transmit awareness messages or a receiver where a VRU equipped with an ITS station may only receive safety warnings or be both a receiver and a transmitter where the VRU device support receiving and transmitting functions. The smartphone is therefore the most appropriate platform for the VRU ITS-station as it can accommodate individual or integrated functions of a heterogeneous system.

Second, modern smartphones already have many strong hardware capabilities and extensive mobile operating systems that facilitate software, internet, and multimedia functionality, alongside core phone and data functions. Almost all the smart devices including, smartphones, smartwatches and tablets have approximately the same internal architecture with differences in size, the number of sensors, and storage capacity. Many smartphones and tablets can enable a wide range of computations of VRU use cases including collision risk calculations and safety warning generation. The periodic broadcast of awareness messages can be achieved through the baseband processor of a smartphone. The smartphone mobile operating system can run personal ITS-S mobile applications and application protocols to enable V2VRU safety interactions. Wireless communication protocols, such as Bluetooth 4, RFID, and Wi-Fi direct, can help VRU detection even without a cellular connection, as tested in previous VRU prototypes and projects. The smartphone GNSS/SBAS positioning unit can directly support many road-level and lane-level VRU safety use cases. The smartphone touchscreen can host a Human Machine Interface (HMI) for awareness warnings and visualization of road users. Besides, third-party location-aware applications, such as Google Maps, can potentially support the prediction of VRU trajectories.

Third, smartphones typically contain many sensors, such as motion sensors, position sensors, environmental sensors. These sensors can provide raw data with high precision and accuracy, making them useful for VRU safety applications. For example, with the use of motion sensors, smartphone-based trajectory prediction algorithms can be implemented to support collision avoidance use cases. The position sensors are useful for VRU navigation and positioning at road-level and lane-level. Sensors embedded in smartphones are therefore useful for implementing functionalities for VRU safety applications such as VRU motion and position detection, thus enabling context-awareness where the current context and the VRU state change are defined for accurate predictions of VRU movements.

Additionally, the use of smartphones for VRU safety applications allows for easy adoption, as by 2021, smartphones are already in the pockets of more than 3.8 billion people globally [67]. Hence, using smartphones as a communication device for VRUs rather than using specialized devices, has clear benefits because it is widespread among road users and has powerful human interfaces and multimedia capabilities [48, 68].

### 4.3 How mobile broadband services can meet the V2VRU communication requirements

In comparison to short-range communication solutions, V2VRU communication based on mobile broadband services (i.e. 4G MBB or 5G MBB) promotes faster market penetration offering financial and implementation benefits due to the ease of integration of the system with portable devices. However, it is worth understanding how mobile broadband services can meet the application requirements mentioned in section 4.1.1:Requirements for V2VRU communication.

The awareness applications require periodic broadcast of basic states of road users. Therefore, the communication network must be able to meet this requirement in order to successfully implement the functionalities of the awareness application. For this purpose, Mobile apps can be designed to exchange data at a specified message size using a smartphone and mobile broadband networks. Currently, 4G MBB has a sufficient data rate to satisfy this requirement. In the coming years, this requirement can easily be fulfilled with 5G MBB services.

The data communication range is another essential requirement of V2VRU communication. The maximum communication range required is specified as 300m for V2M communication [57]. It is also assured that cellular networks, including mobile broadband services, have greater coverage to satisfy this requirement [69]. Similarly, mostly for awareness applications, it is required to communicate data at the maximum frequency of 10Hz satisfying the latency requirement of less than 300ms. In fact, Mobile apps can be programmed for periodic data exchange at the maximum rate of 10Hz using smartphones and mobile broadband services. Experimental evidence is available to demonstrate that mobile broadband networks are capable of transmitting periodic data at a maximum frequency of 10Hz at an average round-trip time (RTT) of less than 300ms [70]. Nevertheless, the data communication performance also depends on the cellular network, communication model, and application protocol. Thus the performance may varies from time to time and location to location. Hence, a complete assessment of performance against the requirement needs extensive field operation testing.

As a solution to the daily increase of road users, the ability to scale the system is the most important feature of a safety application. Accordingly, the V2VRU communication system should handle the increasing number of simultaneously connected devices. A V2VRU system should accommodate urban scenarios with use cases that include 5000 users within the same communication zone (i.e. within a radius up to 300m) [57]. Even with the mobile broadband networks, the required scalability can be achieved by clustering active users and selecting the cluster heads to broadcast the awareness messages. Additionally, by selecting one-to-many or many-to-many communication architecture such as the publish-subscribe communication model, the density of connected users can be increased significantly [70]. Nevertheless, the scalability performance with publish-subscribe models must be assessed by simulations and field operational experiments.

In view of the fact that mobile broadband networks are globally widespread and people are already familiar with the technology, using this technology to introduce safety applications enables applications to be implemented on the application layer using application protocols rather than on the physical layer. Although mobile broadband services such as 4G MBB and 5G MBB are not designed for vehicular networks, the evaluation by David et al. [65] and Araniti et al. [55] and experiments by Liebner et al. [54] show the potential of these technologies for V2VRU communication. This is because of high mobility support, high bit-rates, communication range, and capacity of mobile broadband technologies [35]. More importantly, consideration of this technology for the implementation of V2X communications provides the opportunity for vulnerable road users such as pedestrians, cyclists, and motorcyclists to actively participate in V2X communications and get the benefits of this connected technology. Most significantly, considering this technology to allow V2X communication, vulnerable road users are able to actively engage in V2X communications at a minimal cost and enjoy the benefits of this connected technology. Hence, utilising this technology for implementing V2VRU use cases is more realistic than waiting for C-V2X technology to become popular among users. Nevertheless, it is important to know how to develop such an approach and what technical challenges are associated with this development. Therefore, the next section provides some insights for implementing the VRU safety system using mobile cellular networks.

### 4.4 The proposed V2VRU system framework

The target road users of this system are car/motor vehicle users and vulnerable road users such as pedestrians, cyclists, and motorcyclists. Each road user will carry a communication device (i.e. smartphone). Due to the fact that the existing cellular MBB services provide indirect communication via cellular networks, the road users communicate with each other via a cloud platform that acts as the middleware of the system. The communication between user devices and the cloud server occurs only through cellular wireless technology. Hence, the user devices are connected to the internet via available mobile broadband services.

For V2VRU applications, the main concern is the performance of the application layer connections in terms of latency, reliability, and scalability for connected vehicle applications. The consequent question is, to what degree can the application layer solution support different V2VRU applications? Many internet-based smartphone connections are based on the HTTP client-server architecture. The client-server architecture is a centralized model, in which client requests are sent to the server to receive the information. To overcome the limitations of the client-server architecture and difficulties in the current connected vehicle applications, unidirectional communications based on the publish-subscribe communication paradigm in the application layer is introduced in this system to support the proposed V2VRU data exchanges with MBB under the current 4G-LTE networks and the future 5G eMBB services.

The publish-subscribe communication paradigm has been proposed for distributed real-time applications from previous research studies [71, 72]. This model consists of 3 parties: publisher, subscriber, and middleware, called the publish-subscribe server. The message sender is known as the publisher and the message receiver is known as the subscriber. In the publish-subscribe paradigm, communication between publisher and subscriber takes place according to topics. The publisher defines topics on which they send data and subscribers express their interest in one or more topics to receive data. This method sends immediate notifications of data updates to the subscribers. Receivers need to subscribe to a particular topic only once. The publish-subscribe model can dynamically add and remove participants at any time and thereby, it can be used as a convenient communication system for a large-scale, loosely coupled distributed system such as required by highly dynamic vehicular networks. Due to the multicast communication nature of publish-subscribe communication, once the publisher publishes messages to the publish-subscribe middleware, the message is disseminated to multiple subscribers at once while facilitating scalability to the system. Therefore, safety-related applications can take advantage of the publish-subscribe model to transfer safety notifications among a massive number of road users at the same time. Taking the advantages of the publish-subscribe model, we have adopted it as the primary communication model of our proposed system.

Analyzing system architectures of cellular-based VRU safety systems by Bagheri et al. [35], Lin et al. [37], Zadeh et al. [43], and David et al. [65] a general architecture for the V2VRU communication system based on MBB and smartphone can be presented as Figure 7.

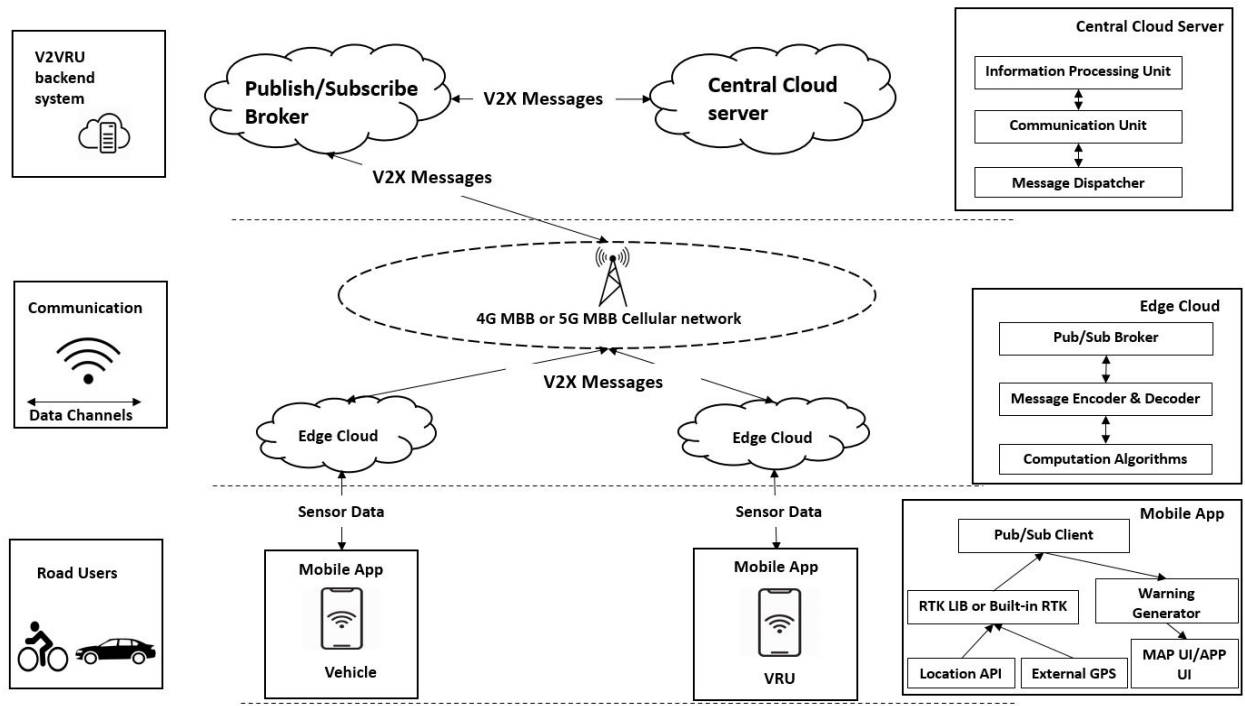

Figure 7: A general system architecture for MBB based V2VRU system

The proposed system architecture consists of a V2VRU Client Application (CA), Edge cloud (EC), and a V2VRU Server Application (SA).

The V2VRU-CA can be installed on smartphones with either 4G MBB or 5G MBB cellular connectivity and is responsible for obtaining sensor data from location API or external GPS receiver to provide user location information. Then the obtained raw GNSS readings are sent to the RTK Lib or built-in RTK to provide GPS corrections. The corrected GPS data are then sent to pub/sub broker installed at the edge cloud via publish/subscribe client. The VRU mobile app is also responsible for generating safety warnings to alert a user of potential collision risks based on the collision risks calculated at the edge cloud.

The edge cloud deployed closer to the end-users is responsible for running computing algorithms to improve system efficiency by reducing system latency. More specifically, this would be a solution to smartphone-related computational complexity and battery usage problems. The sensor data received from the mobile app will be encoded according to the V2VRU message standards using the message encoder. Similarly, the message decoder is responsible for decoding the V2VRU standard messages into mobile app preferred format. The publish/subscribe brokers are installed in both edge clouds and V2VRU SA (i.e. V2VRU backend system) to publish and subscribe to all the V2X messages based on the topics. This reduces the server overhead and improves the efficiency of the system even though the system is scaling up with many users.

The V2VRU-SA runs at central/cloud servers. The V2VRU-CA can access edge clouds deployed closer to V2VRU-SA through the 4G or 5G network. The V2VRU-SA consists of a Communication Unit (CU) that handles C-ITS message communication involving message reception and message transmission among V2VRU-CAs. The Information Processing Unit (IPU) processes the received C-ITS messages from the edge servers, predicts collision risks based on the proximity and user interactions, and takes decisions on disseminating collision warning to V2VRU-CA. The Message Dispatcher (MD) broadcasts warning messages to relevant road users based on the IPU decision.

In this architecture, V2VRU-SA is responsible for performing all the proximity-based computations and publish messages to the pub/sub topics. Then the edge clouds publish the safety messages to the relevant users. V2VRU-CA generates different types of warning notifications (i.e. voice, vibrations, or visual) to alert users of a potential collision. The integration of cloud-based computation architecture with the vehicular network can improve the performance of the V2VRU system by distributing computational functions between the cloud platform and user devices. On the other hand, introducing Mobile Edge Computing (MEC) for V2VRU systems further decreases the communication latency, as this technology increases the computing efficiency by placing computational resources closer to end-users of the network.

### 4.5 Technical challenges for V2VRU communication based on mobile broadband and smartphone

While theories prove that the utilization of smartphones and cellular mobile networks is advantageous for V2VRU communication, several technological challenges hinder the ability of these applications to be implemented in the real world and need to be addressed. Some key technological challenges are outlined below with possible solutions indicated.

*4.5.1 Precise Positioning*

The V2VRU applications with awareness use cases may not require high positioning accuracy since such applications aim to provide road-level or proximity-based awareness. Therefore, situational awareness applications can be implemented with the existing smartphones with the available positioning capability. However, the implementation of collision avoidance applications is challenging. For most of the collision avoidance use cases, precise positioning at an accuracy of 0.5m or higher is required to determine the risk of a potential collision. However, the real-world experiments have repeatedly demonstrated that current smartphones provide positioning uncertainty of 3-10m [32, 34]. The Space-Based Argumentation Systems (SBAS) available in many regions can offer positioning accuracy of 1-2m. However, this is not still sufficiently accurate to meet the 0.5m requirement of V2X applications. Currently GNSS real-time kinematic (RTK) is the only available, widely acceptable solution meeting V2X requirements. However, for connected vehicles and VRU users with smartphones for V2VRU applications, one key challenge for precise positioning is the much higher noise level of the phase measurements due to the low performance of the smartphone built-in GNSS antenna. One possible solution is to use an external antenna. In fact, high precision positioning on smartphones has drawn significant attention from both academia and industries, due to the mass market applications, such as lane-level mapping, traffic monitoring, unique geo-surveying [73]. Recently Google launched the "Google Smartphone Decimeter Challenge" within the GNSS community and provide necessary data sets for competition. The motivation is to encourage the GNSS community to develop high precision GNSS positioning on smartphones. Widely available smartphone decimeter positioning capability will certainly boost V2VRU safety applications.

*4.5.2 Identifying most relevant road users and geomessaging*

Many safety messages of V2VRU applications need to be disseminated to the most relevant road users in a specific geographical area. This is referred to as Geomessaging or GeoCasting. In direct communication since the communication range is small, the users who are within this limited range are considered as most relevant users for receiving the transmitted message.

However, unlike direct communication, cellular technology has the capability of broadcasting messages to multiple users within large proximity. The most common practice of geomessaging in current approaches is circular zone-based proximity communication where the event location is the center of the circle. This involves the transmission of messages to all road users who are within the proximity without filtering the most relevant road users to communicate the particular event based on their location and path of motion. As a result, users receive unnecessary messages that may distract them. Therefore, before communicating the message to nearby users, it is crucial to identify the most potentially at-risk users who may be the victims of the particular event.

The key objective of an efficient geomessaging mechanism is to ensure the successful delivery of messages to most relevant users within the relevant geographical area not only limited to their vicinity but also further filtering them based on the degree of relevancy to the event. Such an effective geomessaging mechanism can be implemented by using IoT protocols that support publish/subscribe communication paradigm [70]. The publish/subscribe paradigm uses message brokers to communicate messages to connected users. The topic-based publish-subscribe communication paradigm allows users to publish or subscribe data to effectively defined topics that support further filtering users based on specific parameters such as position, heading, direction, and orientation.

ETSI EN 302 636-1 standard has described GeoNetworking as another method to achieve geomessaging in mobile networks [74]. GeoNetworking is a network-layer protocol for mobile ad hoc communication based on wireless technology. To achieve geomessaging, GeoNetworking utilizes geographic locations to distribute information and transport data packets. It facilitates the frequent exchange of safety messages at a high rate in the destination area. More importantly, GeoNetworking can send messages to a specific mobile node using its geographical position or multiple mobile nodes in a relevant geographical area.

Another method to achieve geomessaging is dividing geographical map into small squares or set of grids, each with a fixed geographical area and scale. This method allows to separate specific geographical area from the whole map and each road user can be assigned to a grid-based on their location coordinates. Therefore, VRUs and vehicles that are within the particular gird or square can communicate with each other in order to receive the most relevant awareness and collision avoidance warnings. MapTiler [75] is an example tool for dividing a world map into customized tiles.

### 4.5.3 Communication latency

Communication latency is a critical factor in many V2X communication applications. The performance of a V2VRU system highly depends on the communication latency. The end-to-end delay that occurs due to the time taken to communicate data packets from the sender to the receiver and back to the sender is referred to as communication latency. If a V2VRU system could not deliver messages within the required time interval, then the shared data may no longer useful to the receiver for risk calculation or safety warning generation. Hence, the communicated data should be most up-to-date to support both collision avoidance and awareness use cases.

Short-range communication technologies that allow direct communication significantly reduce data communication delay. The latency of ITS-G5/IEEE 802.11p (up to 300m) varies between 1ms and 10ms, compared with 50ms for a high quality of service 4G network. This is mainly because of the propagation time, routing time, and network congestion of the in-direct communication. Therefore, unless using an effective communication protocol and communication mechanism, the centralized nature of MBB services may not be able to support the stringent latency requirement.

The low latency requirement of less than 300ms [57] can be satisfied with the standard uplink/downlink cellular networks by implementing decentralized communication architecture for V2VRU systems rather than adhering to traditional centralized communications.

Therefore, instead of request-response communication model, incorporating decentralized publish/subscribe paradigm [70] that provides asynchronous, scalable and one-to-many or many-to-many anonymous communication, can greatly minimize the end-to-end delay. Many research works have introduced publish-subscribe based edge computing techniques for latency reduction by minimizing both communication and processing delay [76-78]. Data processing closer to the system or at the edge of the system is referred to as edge computing. In edge computing, where the smart devices handle and process data locally is faster than sending data to centralized cloud servers for processing. Therefore, by introducing publish/subscribe based edge computing technology to V2VRU system, both VRU devices and driver devices can process incoming awareness messages in real-time and generate their own warnings to take appropriate actions to avoid collisions based on their basic status data. More importantly, this will reduce bandwidth consumption of the user connections since, in edge computing, VRUs and drivers can operate within the predefined bandwidth limitations of the MBB network [78].

Additionally, Telstra Australia, a leading Australian telecommunications company, has proved that low latency of less than 50 ms can be achieved with existing 4G LTE by optimizing the 4G network through a high-performance Quality of Service (QoS) link [79].

### 4.5.4 Mobile end computation complexity

In a cellular-based V2VRU system, collision risk calculations and safety warning generation algorithms are recommended to run on user end devices in order to increase the performance of the system by reducing server overhead. In fact, when the mobile end performs computations instead of a central server, it should be able to handle computations effectively without any delay and it should not affect the power consumption of the device heavily. Nevertheless, older versions of smartphones may not have the ability of processing complex algorithms due to the computation power of those devices and deprecated technologies. Therefore, to avoid market penetration issues with the V2VRU system, Mobile Edge computing [80, 81] (MEC) can be considered for computation demand. MEC has become an evolving technology that expands the capability of conventional centralized cloud computing to edge closer to end-user devices. Edge computing is proposed for V2X systems along with the topic-based publish/subscribe communication model as described in subsection 4.5.3 Communication latency. The edge computing technology combined with the publish-subscribe model is referred to as the publish-process-subscribe paradigm [76]. By introducing edge computing for V2VRU systems can also meet the latency requirement. Khare et.al [76] suggest a scalable, fog/edge-based broker architecture that operates at the edge of the device to balance data publishing and processing loads. Hence, in the publish-process-subscribe model, data processing and computations are operated at the publish/subscribe brokers deployed at the edge of the system. As a result of this strategy, safety warnings to prevent collisions can be transmitted to subscribed users of the V2VRU system in real-time and the power consumption of consumer devices can be greatly reduced.

### 4.5.5 Interoperability of heterogeneous systems

With the introduction of V2X systems, DSRC and Cellular technology are the two main communication technologies vying for market penetration and acceptance. However, it would also introduce problems of interaction between vehicles that follow only one of the two standards. Enabling interoperability between distinct V2VRU systems is, therefore, a highly important process. Otherwise, it is not possible to exchange safety warnings between such systems, affecting the overall safety of the users. Even within the cellular networks, different stakeholders may be involved in the development of V2X services and may pursue different solutions when offering the same services. It is therefore very challenging to enable interoperability between V2X systems.

One way of ensuring interoperability between systems is to develop and adopt V2X system standards, such as message standards. However, the standardization of V2VRU systems has been initiated most recently. Therefore, in order to enable interoperability between different technologies such as DSRC and Cellular, a proposed solution is to use middleware (i.e. Road Side Unit), equipped with a dual-technology capable server that can transcode from one technology to another [82]. The interoperability among systems of different vendors using the same technology can therefore be achieved by encouraging stakeholders to develop standardized interfaces that follow the same set of standards. The development of V2VRU systems following VRU communication standards can therefore overcome the interoperability issue.

## 5. Conclusion

This paper has discussed the severity of VRU safety issues providing statistical evidence for recent VRU related injuries and fatalities around the world. Driver errors and VRU errors are the key contributing factors for a Vehicle-VRU collision, with reduced visibility and slower reaction time as key contributing factors leading to a collision. Compelling reasons have been made for an assessment of available countermeasures such as roadway designs and engineering solutions, ITS solutions, law enforcement, and safety education. Accordingly, although these countermeasures indeed reduce certain crashes associated with motor vehicles, nonetheless, they cannot substantively address the number of fatalities involving VRUs, as they cannot address the root causes of the problem. Hence, this paper emphasizes the benefits of enabling interconnectivity between vehicles and VRUs to address the key factors leading to a collision. In that context, integrating VRUs with V2X technology is proposed as a promising framework in which to establish better interconnection among motor vehicles and VRUs. The technical and communication requirements for V2VRU use cases have been discussed taking into account the established standards of VRU communication. Following a detailed study of the state-of-the-art technology introduced for V2VRU communication, a general system architecture for smartphones and 4G or 5G MBB based V2VRU communication has been developed. Finally, the challenges of incorporating VRUs into the V2X context using current cellular networks are discussed with a range of suggestions to pursue the implementation of VRU safety in a new direction.

### Acknowledgment

The first author acknowledges the scholarship support from the Queensland University of Technology.

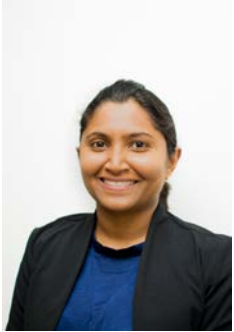

NISHANTHI DASANAYAKA received the B.Sc. degree in Computer Science from University of Kelaniya, Sri Lanka in 2015. She is currently pursuing her Ph.D. in Computer Science at Queensland University of Technology, Brisbane, Australia. Her current research focuses on Intelligent Transportation Systems, effective use of Vehicle-to-Everything (V2X) communication, Vehicle-to-VRU (V2VRU) communication and Vehicle-to-Infrastructure (V2I) communication for road safety, and publish-subscribe based IoT protocols.

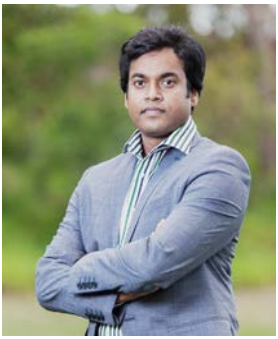

KHONDOKAR FIDA HASAN (S'13, M'18) received a PhD from the School of Electrical Engineering and Computer Science, Queensland University of Technology (QUT), Australia. He has awarded the Fellow of HEA, UK, for the excellence of his research and teaching practice in higher academia. His current research interests include emerging Network Technologies and Intelligent Systems, Road Safety and Intelligent Transportation Systems, and Blockchain application in the Internet of Things.

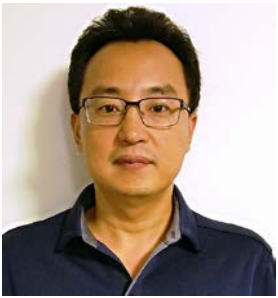

CHARLES WANG is a Research Fellow in the Science and Engineering Faculty of Queensland University of Technology (QUT), Australia. His major research interests are in GNSS data processing, GNSS orbit estimation and precise positioning for emerging applications such as Corporate Intelligent Transport Systems and Internet of Things.

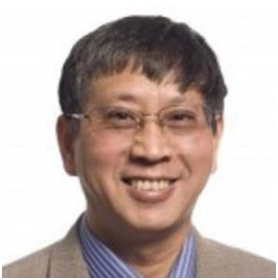

YANMING FENG is currently a professor in Data Science at the School of Computer Science, Queensland University of Technology, Australia. His active research interests include orbit determination, wide area GNSS positioning, multiple-frequency GNSS data processing, integrity determination and vehicle communications networks. His recent work focuses on time synchronisation of vehicles networks, vehicle positioning and Internet of Things for industry applications. Yanming Feng held BSC, MSC and PhD degrees in geodesy and spatial science from Wuhan University, China.